\documentclass[aps,twocolumn,superscriptaddress,amsmath,longbibliography,nofootinbib]{revtex4-1}
\usepackage{graphicx,color,hyperref,xcolor}
\usepackage{soul}
\usepackage{amsfonts}
\usepackage{amssymb}
\usepackage{amsmath}
\usepackage{amsthm}
\usepackage{bm}
\usepackage{braket}
\usepackage{multirow}
\usepackage{array}

\usepackage[makeroom]{cancel}

\makeatletter
\newcommand*{\rom}[1]{\expandafter\@slowromancap\romannumeral #1@}
\makeatother

\begin{document}

\title{Microscopic Investigation of rf Vortex Nucleation in Nb\textsubscript{3}Sn Films Using a Near-Field Magnetic Microwave Microscope}

\author{Chung-Yang Wang}
\affiliation{Quantum Materials Center, Department of Physics, University of Maryland, College Park, Maryland 20742, USA}

\author{Zeming Sun}
\affiliation{Department of Physics, Cornell University, Ithaca, NY, USA}

\author{Thomas Oseroff}
\affiliation{Department of Physics, Cornell University, Ithaca, NY, USA}

\author{Matthias U. Liepe}
\affiliation{Department of Physics, Cornell University, Ithaca, NY, USA}

\author{Steven M. Anlage}
\affiliation{Quantum Materials Center, Department of Physics, University of Maryland, College Park, Maryland 20742, USA}

\begin{abstract}

We use a near-field magnetic microwave microscope to investigate and compare rf vortex nucleation in two superconducting radio-frequency (SRF)-quality Nb\textsubscript{3}Sn films fabricated by different methods: a conventional vapor-diffused film and an electrochemically plated film followed by thermal annealing, both of which are deposited on Nb substrates. The microscope applies a localized rf magnetic field to the sample surface and measures the resulting third-harmonic response $P_\mathrm{3f}$, which is particularly sensitive to rf vortex nucleation triggered by surface defects. Both Nb\textsubscript{3}Sn films exhibit nontrivial $P_\mathrm{3f}(T)$ structures below 7~K that display the key signatures associated with rf vortex nucleation at local defects. The electrochemical film additionally shows multiple $P_\mathrm{3f}(T)$ structures between 14~K and 16~K that are absent in the vapor-diffused sample. Our results highlight the influence of fabrication method on rf vortex penetration properties and demonstrate the utility of third-harmonic response as a local diagnostic tool for surface defects in Nb\textsubscript{3}Sn films.

\end{abstract}

\maketitle



\section{Introduction}
\label{sec:Introduction}

As a promising material for next-generation superconducting radio-frequency (SRF) cavities \cite{Pad17,AnlageIEEEJMicro21}, Nb\textsubscript{3}Sn offers significant performance advantages over conventional niobium. Its high critical temperature ($T_\mathrm{c} \approx 18$~K), large superconducting gap, and high superheating field support operation at higher accelerating gradients, offer lower surface resistance, and reduced cryogenic load~\cite{Kneisel77,posen2015proof,posen2015radio,posen2017nb3sn}. In particular, the ability to operate SRF cavities near 4~K using compact commercial cryocoolers—rather than complex helium cryomodules—promises to substantially lower system cost and operational complexity~\cite{posen2017nb3sn}. These advantages have driven extensive research into the development of high-quality Nb\textsubscript{3}Sn films optimized for SRF performance~\cite{Becker15,ilyina2019development,posen2015proof,Sun2021Toward,Sun2023Smooth,Howard2023Thermal}.

Despite these advantages, Nb\textsubscript{3}Sn films—particularly those fabricated by the conventional vapor diffusion process~\cite{posen2017nb3sn}—face several materials challenges that limit their rf performance. Key issues include excessive surface roughness~\cite{porter2016surface,porter2019progress,posen2021advances,pack2020vortex}, local stoichiometric deviations~\cite{trenikhina2017performance,carlson2021analysis}, Sn segregation at grain boundaries~\cite{lee2020grain,carlson2021analysis}, and impurity incorporation~\cite{Sun2023Surface}, all of which can degrade the superconducting properties and contribute to rf losses~\cite{carlson2021analysis}. Surface roughness can locally enhance rf magnetic fields, triggering premature vortex entry and quench~\cite{porter2016surface,porter2019progress,posen2021advances,pack2020vortex}. Stoichiometric inhomogeneities, including both Sn-deficient and Sn-rich regions, introduce spatial variations in $T_\mathrm{c}$, energy gap $\Delta$, and enhance rf dissipation~\cite{Becker15,trenikhina2017performance,LeeSnDeficient19,carlson2021analysis}. Grain boundaries are also recognized as potential weak points; recent studies have revealed significant Sn segregation and local suppression of superconductivity at grain boundaries~\cite{lee2020grain}, which may act as favorable sites for rf vortex nucleation under strong microwave fields~\cite{carlson2021analysis}. 

Overcoming these limitations requires a detailed understanding of local superconducting behavior and the continued development of fabrication methods that improve surface quality. Here, ‘surface properties’ include both chemical (stoichiometry and impurity levels) and physical (surface roughness, grain shape/size, and grain-boundary geometry and population). The deposition route and its associated phase transformations govern grain growth and the resulting surface properties. For example, electrochemical Sn pre-deposition followed by thermal annealing forms a continuous nucleation seed layer, whereas vapor diffusion relies on adsorption of Sn vapor at initial discrete sites until a nucleation layer coalesces. As a result, electrochemically prepared Nb\textsubscript{3}Sn films exhibit improved stoichiometry and reduced roughness ($R_a \approx 60$~nm) compared to typical vapor-diffused films ($R_a \approx 300$~nm),~\cite{Sun2021Toward,Sun2023Smooth} albeit with smaller average grains and more heterogeneous local surface features due to the higher density of nucleation sites. Technologically, vapor diffusion uses comparatively simpler tooling but allows less local control of flux than a pre-deposited Sn layer; electrochemistry enables sufficient local Sn supply but is a specialized process with bath management and operator-dependent variability. These process trade-offs also impact the surface properties, and therefore, nonlinear RF behavior.

As discussed above, rf vortex nucleation arising from surface roughness, stoichiometric inhomogeneity, and Sn segregation at grain boundaries can contribute to losses and limit the rf performance of Nb\textsubscript{3}Sn. These mechanisms are inherently local, often originating from nanoscale surface defects that are not easily detected by macroscopic measurement techniques. It is therefore beneficial to adopt a microscopic approach that enables local investigation of rf vortex nucleation due to surface defects. Motivated by this need, we have developed a near-field magnetic microwave microscope~\cite{lee2000magnetic,lee2003Study,lee2003spatially,lee2005doping,lee2005microwave,mircea2009phase,tai2011nonlinear,tai2012nanoscale,tai2014modeling,tai2014near,tai2015nanoscale,oripov2019high,wang2024microscopic,Wang2024Near,wang2025microwave} that applies a localized rf magnetic field to a confined region of the sample and measures the locally-generated third-harmonic response.

The third-harmonic signal $P_\mathrm{3f}$ is particularly sensitive to rf vortex nucleation: when the applied rf magnetic field exceeds the local vortex penetration field, transient rf semi-loop vortices nucleate and produce a strong nonlinear response at the third harmonic~\cite{oripov2020time,wang2024microscopic}. In a previous study on SRF-quality Nb/Cu films~\cite{wang2024microscopic}, this technique successfully identified surface defects capable of nucleating rf vortices by analyzing the temperature and input power dependence of $P_\mathrm{3f}$.

In this work, we apply our near-field magnetic microwave microscope \cite{AnlageNFMM2001,AnlageSPMBook2007} to measure third-harmonic response and investigate rf vortex nucleation in two SRF-quality Nb\textsubscript{3}Sn films prepared by different fabrication methods at Cornell University: a traditional vapor-diffused film and an electrochemically plated film followed by thermal annealing~\cite{Sun2021Toward,Sun2023Smooth}. By analyzing the temperature and input power dependence of $P_\mathrm{3f}$, we identify nontrivial features consistent with rf vortex nucleation by surface defects. A comparison between the two films provides insight into how surface properties resulting from different fabrication methods influence the nature and distribution of these nonlinear responses.

The structure of this paper is as follows. Section~\ref{sec:Setup} describes the experimental setup of the near-field magnetic microwave microscope used to probe local nonlinear electrodynamics. Section~\ref{sec:ExperimentalResults} presents the third-harmonic response data obtained from the two Nb\textsubscript{3}Sn films, along with a detailed analysis focusing on signatures of surface defects. Section~\ref{sec:Discussion} compares the results from both samples and discusses their implications in the context of rf vortex nucleation. Finally, Section~\ref{sec:Conclusion} summarizes the key findings and their relevance to understanding and optimizing Nb\textsubscript{3}Sn films for SRF applications.


\section{Experimental setup}
\label{sec:Setup}

The experimental setup of the near-field magnetic microwave microscope used in this study is identical to that described in previous works~\cite{oripov2019high,Wang2024Near,wang2024microscopic}. For completeness, a brief overview of the microscope and its operating principles is provided in this section.

The core component of the microwave microscope is a magnetic writer probe, originally designed for conventional hard-disk drives and provided by Seagate Technology. This probe is capable of generating a highly localized and intense rf magnetic field on the surface of the superconductor.

The measurement process begins with the production of a microwave signal $P_\mathrm{rf}\mathrm{sin}^{2}(\omega t)$ at frequency $f=\omega/2\pi$ by a microwave source. This signal is transmitted to the magnetic writer probe inside the cryostat via a series of coaxial cables and the built-in transmission line of the probe. The probe then produces a localized rf magnetic field $B_\mathrm{rf}\mathrm{sin}(\omega t)$, which is applied to the surface of the superconducting sample. In response, the sample generates a time-dependent screening current on and near the surface in an effort to maintain the Meissner state. The magnetic field associated with this screening current is then coupled back to the same probe, creating a return propagating signal on the transmission line. This return signal contains multiple harmonic components, from which the third-harmonic component $P_\mathrm{3f}\mathrm{sin}^{2}(3 \omega t)$ is isolated using appropriate filtering. The power of this third-harmonic signal is then extracted and measured by a spectrum analyzer at room temperature.

Measurements are performed with a variety of fixed input frequencies ranging from 1.1~GHz to 2.2~GHz, while systematically varying the sample temperature and applied rf field amplitude. The base temperature for a sample in the cryostat is around 3.6 K. 

During microwave measurements, the probe is in contact with the sample surface. However, the surfaces of the probe and the sample are not perfectly flat, resulting in a finite probe-sample separation $h$, which influences both the field of view of the probe and the peak rf magnetic field experienced by the sample $B_\mathrm{pk}$. The separation $h$ is estimated to be less than 1~$\mu$m, and for an input microwave power of 0~dBm, $B_\mathrm{pk}$ is estimated to lie within the range of several tens of millitesla~\cite{oripov2019high,Wang2024Near,wang2024microscopic,wang2025microwave}. The field of view of the probe ranges from sub-micron to micron scale, depending on the probe-sample separation and the specific signal being analyzed~\cite{wang2024microscopic,wang2025microwave}.

No external DC magnetic field is applied during the measurements. The residual DC field near the sample at low temperatures is measured to be approximately 35~$\mu$T using a cryogenic three-axis magnetometer. The corresponding average separation between trapped flux quanta, given by $\sqrt{\Phi_{0}/B}$, is approximately 7.7~$\mu$m, which is significantly larger than the field of view of the microscope probe. Therefore, even if the entire 35~$\mu$T residual DC field were to become trapped in the sample, the probability of having a DC vortex directly beneath the probe is low.  We note that prior work has examined the microwave flux flow resistivity and depinning frequency of Nb\textsubscript{3}Sn bulk material in large DC magnetic fields \cite{AlimentiBulkDePin21}.

Background nonlinear measurements have been performed with the same type of probe in contact with a copper surface as a function of temperature \cite{wang2024microscopic}. These measurements show no signs of background nonlinearity arising from the wiring or magnetic properties of the probe for the input power levels used here.


\section{Experimental Results}
\label{sec:ExperimentalResults}

In this work, we investigate two Nb\textsubscript{3}Sn films: one deposited using the traditional vapor diffusion process (vapor-diffused Nb\textsubscript{3}Sn film) and the other synthesized through an electrochemical plating process followed by thermal annealing (electrochemical Nb\textsubscript{3}Sn film). Both films were fabricated at Cornell University. Details on the deposition methods, surface roughness, local stoichiometry, and other characteristics of these films can be found in Refs.~\cite{Sun2021Toward,Sun2023Smooth,Sun2023Surface,ZemingThermo23}.

For the measurements on the vapor-diffused Nb\textsubscript{3}Sn film, the microscope probe was mounted on a cryogenic XYZ positioner and operated in a scanning probe microscope configuration. In contrast, for the measurements on the electrochemical Nb\textsubscript{3}Sn film, the probe was fixed in place using a copper clamp and pressed directly against the sample surface to achieve a reduced probe-sample separation.

\subsection{Sample Information}
\label{sec:SampleInformation}

\begin{figure}
\includegraphics[width=0.45\textwidth]{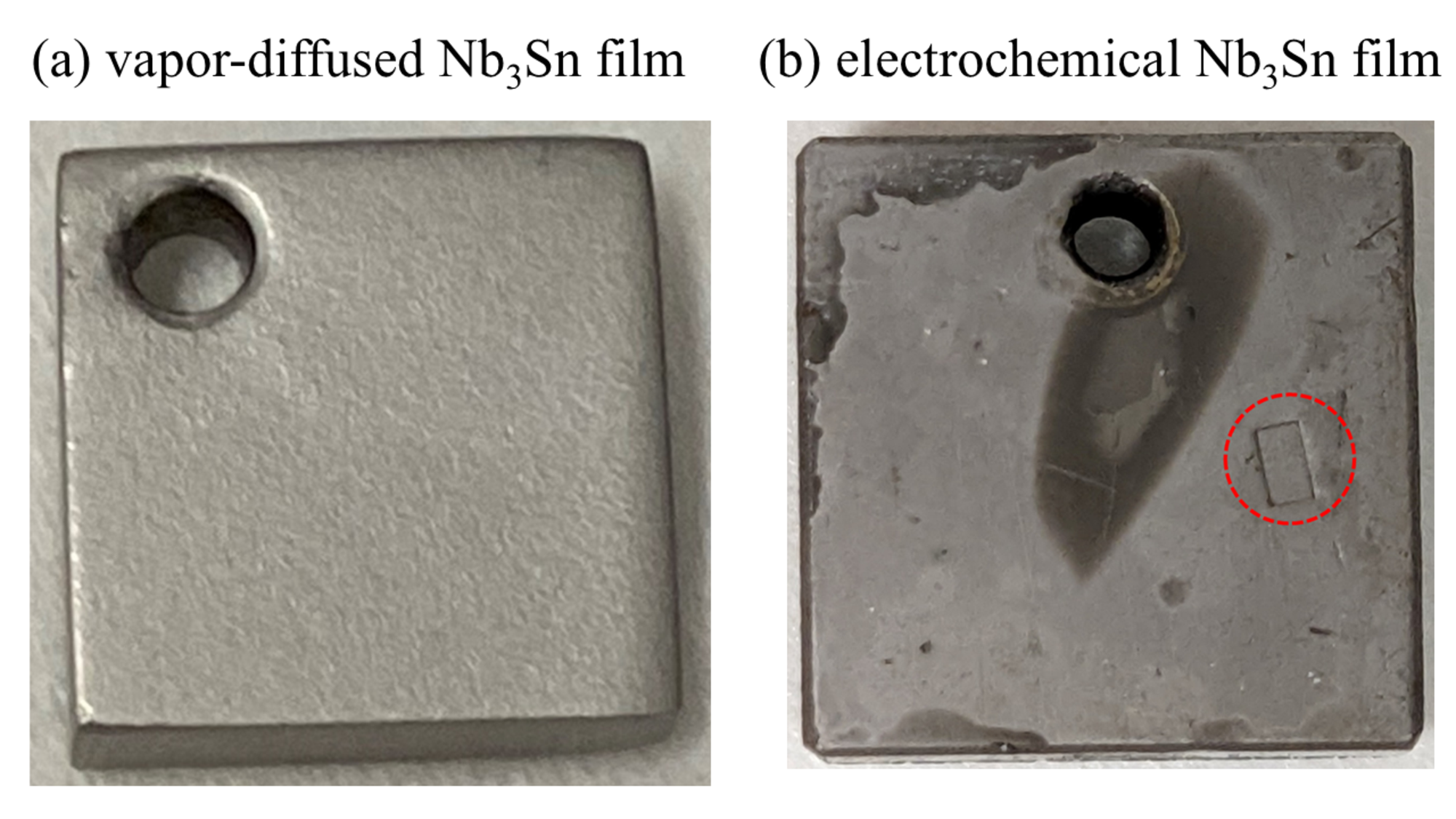}
\caption{\label{fig:samplephoto}Photographs of the two Nb\textsubscript{3}Sn films used in this study, both prepared at Cornell University. (a) Vapor-diffused Nb\textsubscript{3}Sn film. (b) Electrochemical Nb\textsubscript{3}Sn film. Each sample has lateral dimensions of 1~cm~$\times$~1~cm. The red dashed circle in (b) highlights a scar on the surface caused by an aggressive probe contact during a later measurement, which did not affect the data presented in the main text.}
\end{figure}

Figure~\ref{fig:samplephoto} shows the two Nb\textsubscript{3}Sn films studied: a vapor-diffused film (Fig.~\ref{fig:samplephoto}(a), nominal 2.0~$\mu$m thick) and an electrochemically deposited film (Fig.~\ref{fig:samplephoto}(b), 2.4~$\mu$m thick). Both were prepared on the same grade of high-purity Nb (RRR $\approx$ 300), diced from 3-mm thick plates into 1~cm~$\times$~1~cm coupons. Sn electrodeposition employed a three-electrode system (Pt counter; SCE reference) in an aqueous seed-free SnCl\textsubscript{2} / ammonium citrate bath. Sn pre-deposits were converted to Nb\textsubscript{3}Sn by high-vacuum annealing ($10^{-6}$ Torr): 500\textdegree C for 5 h (nucleation) followed by 1100\textdegree C for 3 h (alloying), then furnace cooling. Vapor-diffused samples used the same thermal schedule with a separate Sn source providing a controlled Sn-vapor flux at tuned vapor pressure \cite{Sun2023Smooth}.


Third-harmonic response measurements were performed on the electrochemical Nb\textsubscript{3}Sn film across multiple sessions. The data presented in the main text were obtained during early-stage measurements, prior to any observable surface damage. In a subsequent session, the microscope probe was pressed against the sample with a substantial force, leaving a visible scar on the surface, as indicated by the red dashed circle in Fig.~\ref{fig:samplephoto}(b). The results obtained after this incident are presented in Appendix~\ref{sec:DamagedResult}.

\subsection{Experimental Results for the Vapor-Diffused Nb\textsubscript{3}Sn Film}
\label{sec:DataVaporDiffused}

\begin{figure}
\includegraphics[width=0.45\textwidth]{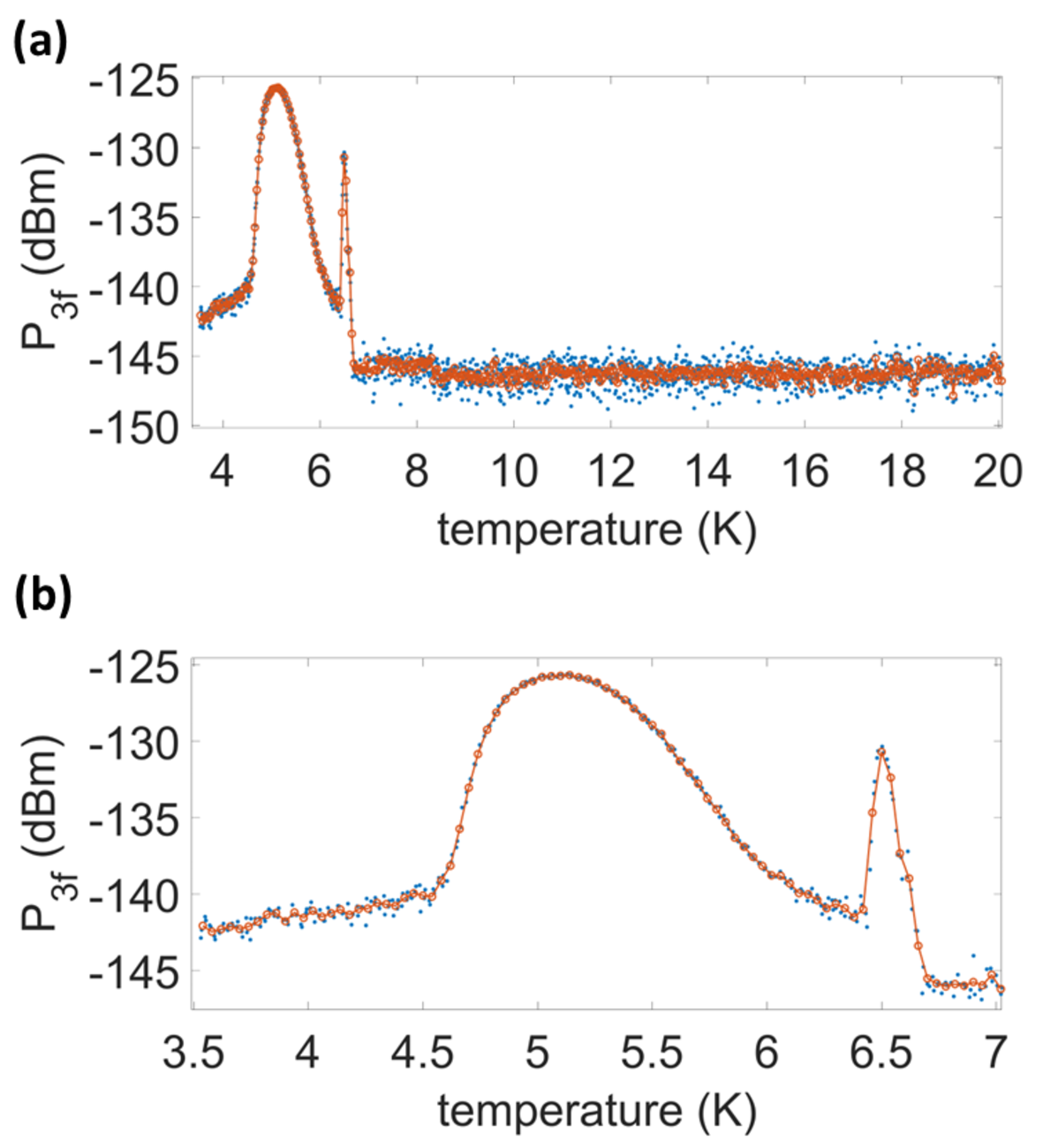}
\caption{\label{fig:vapordiffuseTsweepsingleP}Representative data for $P_\mathrm{3f}$ as a function of temperature for the vapor-diffused Nb\textsubscript{3}Sn film. The input frequency is 2.08 GHz and the input power is -8 dBm. The blue dots are the raw data, and the red curve is the $P_\mathrm{3f}$ averaged over 0.04 K range bins. (a) Full temperature range from 3.6 K to 20 K. (b) Expanded view of the low-temperature region below 7 K.}
\end{figure}

Figure~\ref{fig:vapordiffuseTsweepsingleP} shows representative third-harmonic response data as a function of temperature, $P_\mathrm{3f}(T)$, measured at a fixed location on the vapor-diffused Nb\textsubscript{3}Sn film. The measurement protocol proceeds as follows: the sample is first warmed to 20~K, above its superconducting transition temperature $T_\mathrm{c}$ ($\approx$ 18~K), and then the microwave source is activated with a fixed input frequency ($f=\omega/{2\pi}$=2.08 GHz) and input power ($P$=-8 dBm). The third-harmonic response $P_\mathrm{3f}$ is measured as the sample is gradually cooled down to 3.6~K.

The measured $P_\mathrm{3f}(T)$ signal in Fig.~\ref{fig:vapordiffuseTsweepsingleP} contains contributions from both the sample and a temperature-independent background originating from the microscope probe. The magnetic writer probe used as the microscope probe exhibits intrinsic nonlinearity, which remains constant across temperature. This probe background dominates the signal above 18~K. A significant increase in $P_\mathrm{3f}$ is observed below 6.7~K, as shown in Fig.~\ref{fig:vapordiffuseTsweepsingleP}(b). We define this onset temperature of strong third-harmonic generation as the $P_\mathrm{3f}$ transition temperature, denoted $T_\mathrm{c}^\mathrm{P_{3f}}$. The emergence of enhanced nonlinearity below $T_\mathrm{c}^\mathrm{P_{3f}}$ indicates the presence of additional mechanisms—likely extrinsic in origin—associated with material capable of generating strong nonlinear microwave response.

One may notice that the intrinsic Nb\textsubscript{3}Sn response near its bulk superconducting transition around 18~K is absent in Fig.~\ref{fig:vapordiffuseTsweepsingleP}. This is because our local $P_\mathrm{3f}$ measurements are primarily sensitive to surface defects, rather than the intrinsic nonlinear response of the superconductor \cite{wang2024microscopic}. Since the primary objective of this work is to investigate the rf properties of surface defects, the strong third-harmonic response observed below 6.7~K, as shown in Fig.~\ref{fig:vapordiffuseTsweepsingleP}, forms the central focus of the following analysis.

\begin{table}
\begin{center}
\begin{tabular}{ | m{1.5cm} | m{6.5cm}| }
 \hline
 Feature 1 & $P_\mathrm{3f}(T)$ onsets below $T_\mathrm{c}$ of Nb\textsubscript{3}Sn and shows a bell-shaped structure. \\ 
 \hline
 Feature 2 & $P_\mathrm{3f}(T)$ maximum increases for a stronger driving rf field amplitude. \\ 
 \hline
 Feature 3 & $P_\mathrm{3f}(T)$ maximum shows up at a lower temperature for a stronger driving rf field amplitude. \\ 
 \hline
 Feature 4 & $P_\mathrm{3f}(T)$ low-temperature onset shows up at a lower temperature for a stronger driving rf field amplitude. \\ 
 \hline
\end{tabular}
\caption{The four key features of $P_\mathrm{3f}(T)$ due to rf semi-loop vortex nucleation associated with surface defects.}
\label{tbl:FourKeyFeatures}
\end{center}
\end{table}

As presented in previous work \cite{wang2024microscopic}, the microwave microscope can locally nucleate rf semi-loop vortices \cite{GurSemiLoop08} at surface defects beneath the probe, producing a third-harmonic response $P_\mathrm{3f}(T)$ that exhibits four characteristic features. In that study, time-dependent Ginzburg-Landau (TDGL) simulations were also performed using a phenomenological grain boundary toy model, where grain boundaries were modeled as regions containing low-$T_\mathrm{c}$ impurity phases. Notably, both the experimental measurements and the TDGL simulations revealed the same four robust features in $P_\mathrm{3f}(T)$, indicative of rf semi-loop vortex nucleation at grain boundaries. These four features are summarized in Table~\ref{tbl:FourKeyFeatures} and serve as diagnostic criteria in the following analysis for identifying defect-induced rf semi-loop vortex nucleation in the Nb\textsubscript{3}Sn films.

\begin{figure}
\includegraphics[width=0.45\textwidth]{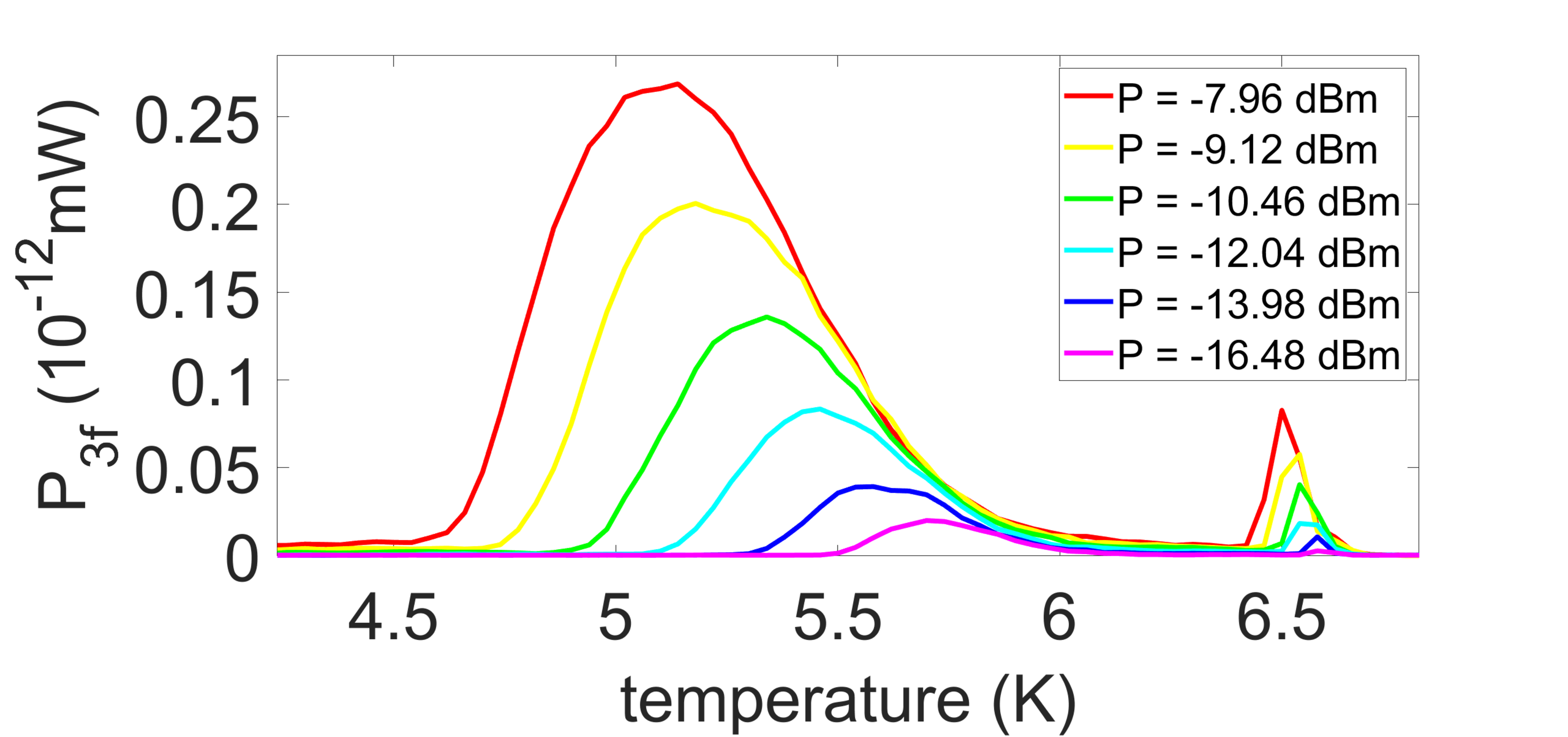}
\caption{\label{fig:vapordiffuseTsweepmultipleP}Measured $P_\mathrm{3f}(T)$ on a linear power scale for the vapor-diffused Nb\textsubscript{3}Sn film at an input frequency of 2.08~GHz, with input power decreasing from strong (red) to weak (purple). Compared to the raw data shown in Fig.~\ref{fig:vapordiffuseTsweepsingleP}, here the probe background is subtracted. The input power levels are not evenly spaced in power but rather in rf field amplitude. Note that the input power is proportional to the square of the rf field amplitude.}
\end{figure}

To further investigate the nature of the third-harmonic response of the vapor-diffused Nb\textsubscript{3}Sn film below 6.7~K, as shown in Fig.~\ref{fig:vapordiffuseTsweepsingleP}, $P_\mathrm{3f}(T)$ measurements are performed at a series of input powers $P$, corresponding to different applied rf field amplitudes. For each measurement, the sample is first warmed to 20~K and then gradually cooled to 3.6~K while subjected to a fixed-frequency rf field (2.08~GHz) at a constant input power. To isolate the sample response, the probe background is estimated by averaging the $P_\mathrm{3f}(T)$ magnitude between 19~K and 20~K and is subtracted from the total signal. This procedure is repeated for six different input powers. The resulting background-subtracted $P_\mathrm{3f}(T)$ curves at a fixed location on the vapor-diffused Nb\textsubscript{3}Sn film are shown on a linear power scale in Fig.~\ref{fig:vapordiffuseTsweepmultipleP}. The input power levels are chosen to span uniform steps in rf field amplitude, rather than in power. Note that the input power is proportional to the square of the rf field amplitude.

Two distinct nontrivial structures are observed in Fig.~\ref{fig:vapordiffuseTsweepmultipleP}, corresponding to $P_\mathrm{3f}(T)$ onsets (upon cooling) at $T_\mathrm{c}^\mathrm{P_{3f}} = 6.2$~K and $T_\mathrm{c}^\mathrm{P_{3f}} = 6.7$~K, respectively. The presence of two separate $T_\mathrm{c}^\mathrm{P_{3f}}$ values suggests that the measured $P_\mathrm{3f}(T)$ is the superposition of two defect signals. Both $P_\mathrm{3f}(T)$ structures exhibit the four key characteristics outlined in Table~\ref{tbl:FourKeyFeatures}, indicating that they both likely originate from rf semi-loop vortex nucleation associated with surface defects. These results indicate that the vapor-diffused Nb\textsubscript{3}Sn film contains at least two distinct surface defects, each capable of nucleating rf vortices but with different associated $P_\mathrm{3f}$ transition temperatures due to local variation in superconducting properties.

\begin{figure}
\includegraphics[width=0.45\textwidth]{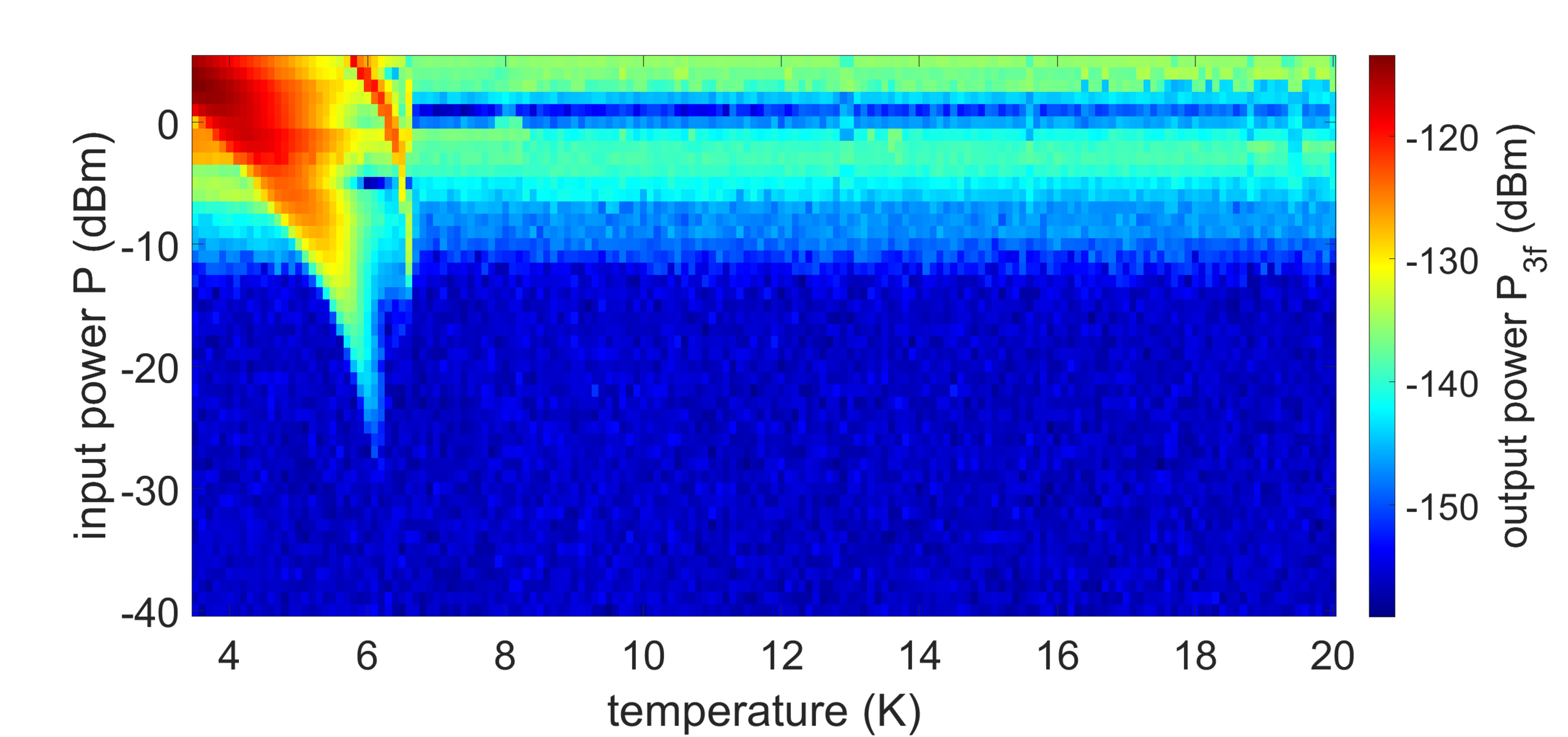}
\caption{\label{fig:vapordiffuseTPsweep}Color map of the measured third-harmonic power $P_\mathrm{3f}$ (in dBm, log scale) as a function of temperature $T$ and input power $P$ for the vapor-diffused Nb\textsubscript{3}Sn film. The input frequency is 2.08 GHz. The input-power-dependent probe background is not subtracted.}
\end{figure}

To complement the $P_\mathrm{3f}(T)$ measurements performed over a limited input power range, we also conducted a broader input power sweep, as shown in Fig.~\ref{fig:vapordiffuseTPsweep}. This measurement spans temperatures from 3.6~K to 20~K, exceeding the intrinsic $T_\mathrm{c}$ of Nb\textsubscript{3}Sn, and input powers from $-40$~dBm to $+5$~dBm. Across this expanded parameter space, no additional nontrivial $P_\mathrm{3f}(T)$ structures are observed beyond the two features previously identified in Fig.~\ref{fig:vapordiffuseTsweepmultipleP}.

\subsection{Experimental Results for the Electrochemical Nb\textsubscript{3}Sn Film}
\label{sec:DataSnPlated}

\begin{figure}
\includegraphics[width=0.45\textwidth]{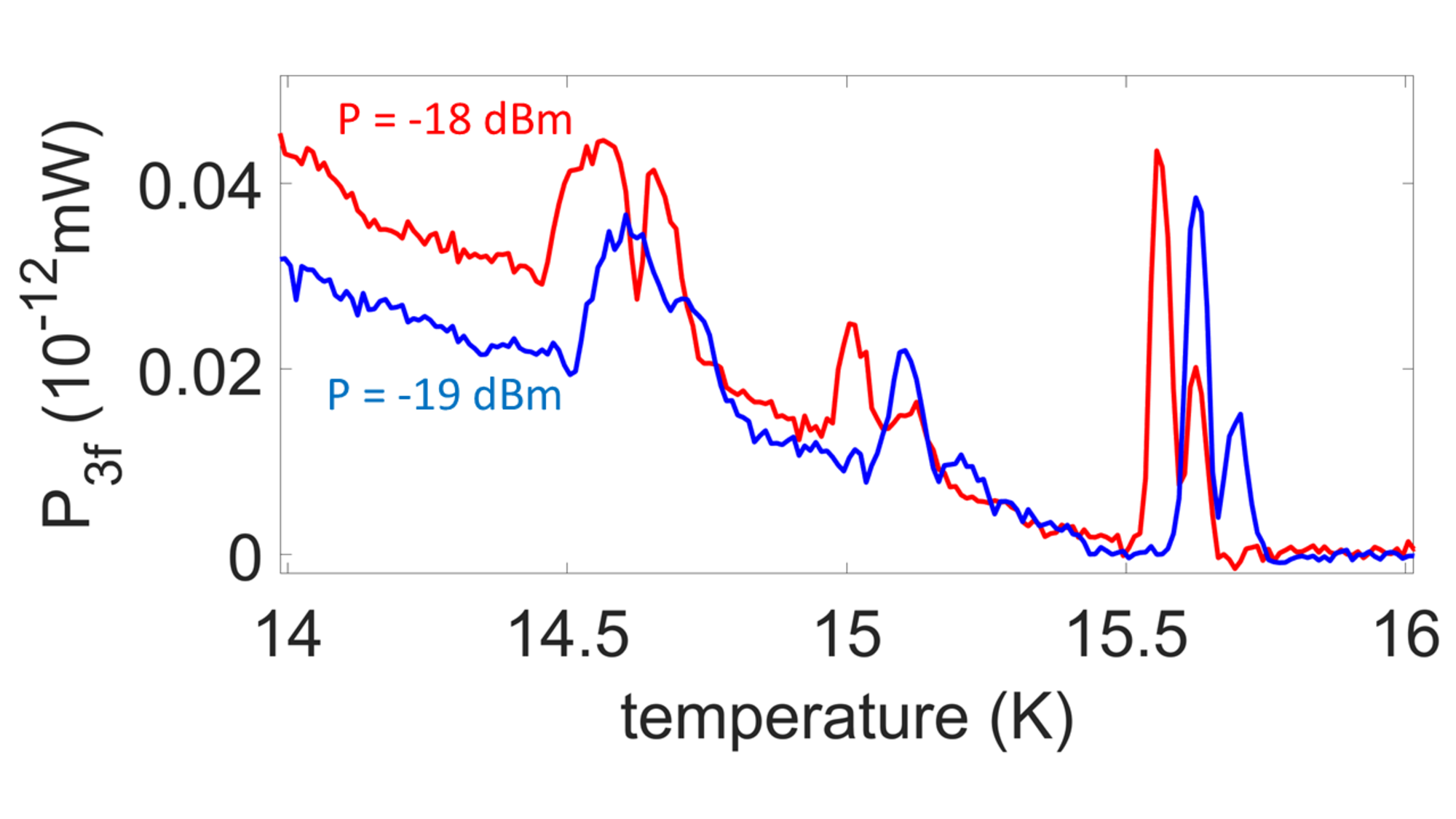}
\caption{\label{fig:SnplatedTsweepmultipleP}Measured linear power scale $P_\mathrm{3f}(T)$ for the electrochemical Nb\textsubscript{3}Sn film at an input frequency of 1.21 GHz. Data are shown for two different input powers: $P = -18$ dBm (red) and $P = -19$ dBm (blue). The probe background has been subtracted.}
\end{figure}

The temperature and input power dependence of $P_\mathrm{3f}$ in the electrochemical Nb\textsubscript{3}Sn film are studied using the same measurement protocol as for the vapor-diffused film. Figure~\ref{fig:SnplatedTsweepmultipleP} shows the third-harmonic response $P_\mathrm{3f}(T)$ on a linear power scale, measured at a fixed location on the electrochemical Nb\textsubscript{3}Sn film for two different input powers. The input frequency is 1.21~GHz, and the probe background has been subtracted from the data. While $P_\mathrm{3f}(T)$ is recorded over the full temperature range from 20~K to 3.6~K, Fig.~\ref{fig:SnplatedTsweepmultipleP} highlights the nontrivial $P_\mathrm{3f}(T)$ structures that emerge between 14~K and 16~K.

Three distinct nontrivial $P_\mathrm{3f}(T)$ structures are observed in Fig.~\ref{fig:SnplatedTsweepmultipleP}: one around 15.6~K, another around 15.1~K, and a third around 14.6~K. Each of these $P_\mathrm{3f}(T)$ structures exhibits the four key features outlined in Table~\ref{tbl:FourKeyFeatures}, indicating that they are likely associated with rf vortex nucleation by surface defects. The presence of multiple $P_\mathrm{3f}$ transitions at different temperatures suggests that the electrochemical Nb\textsubscript{3}Sn film contains several surface defects with varying local superconducting properties.

\begin{figure}
\includegraphics[width=0.45\textwidth]{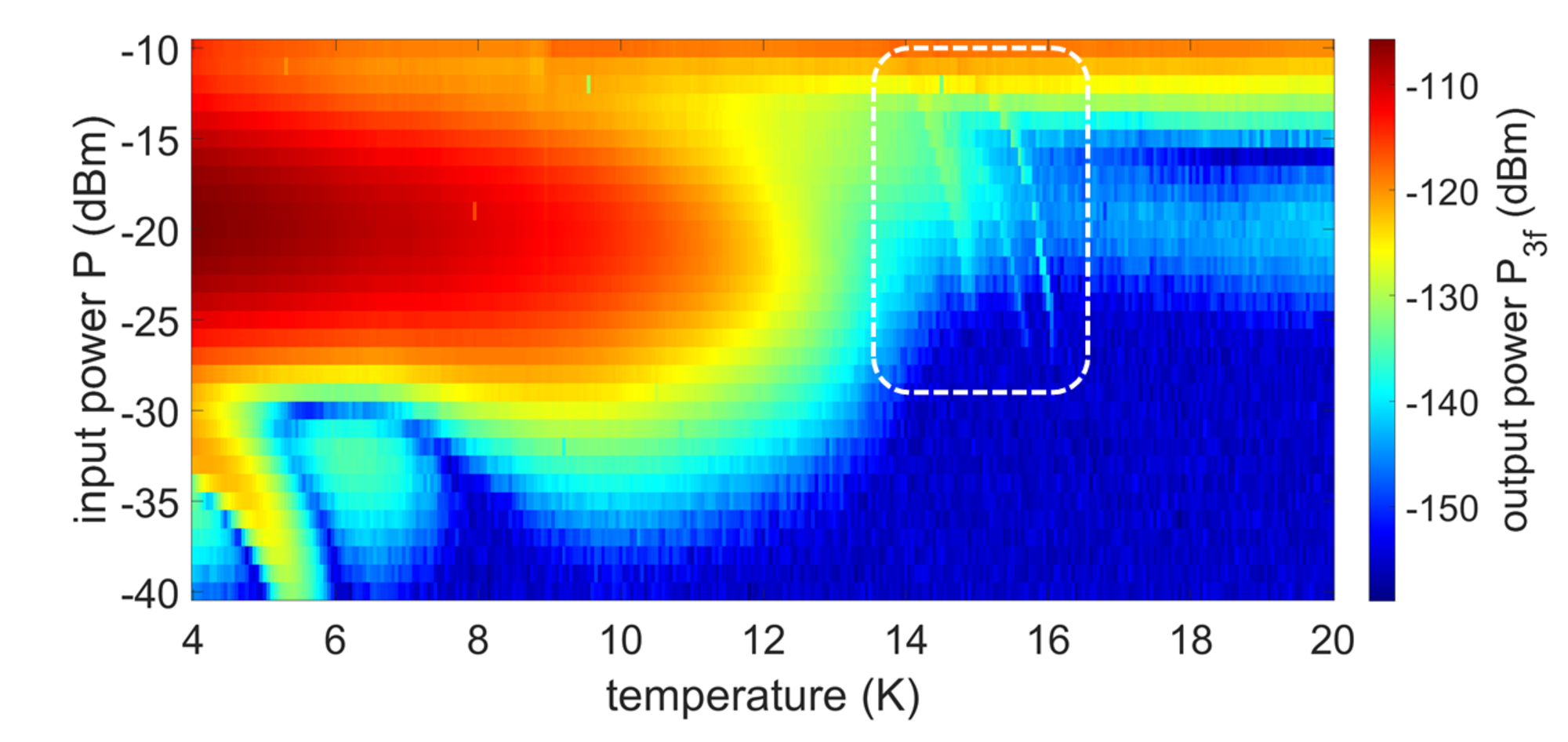}
\caption{\label{fig:SnplatedTPsweep}Color map of the measured log power scale $P_\mathrm{3f}$ as a function of temperature $T$ and input power $P$ for the electrochemical Nb\textsubscript{3}Sn film. The input frequency is 1.21 GHz. The input-power-dependent probe background is not subtracted. The white dashed box highlights the region containing the three nontrivial $P_\mathrm{3f}(T)$ structures, whose details are shown in Fig.~\ref{fig:SnplatedTsweepmultipleP}.}
\end{figure}

In addition to the $P_\mathrm{3f}(T)$ measurements over a limited input power range, we also perform a broader scan across both temperature and input power, as shown in Fig.~\ref{fig:SnplatedTPsweep}. This figure presents the measured third-harmonic power $P_\mathrm{3f}$ (in dBm, color scale) as a function of temperature $T$ (horizontal axis) and input power $P$ (vertical axis) for the electrochemical Nb\textsubscript{3}Sn film. The input frequency is 1.21~GHz, and the input-power-dependent probe background is not subtracted. The measurement spans temperatures from 4~K to 20~K, and input powers from $-40$~dBm to $-10$~dBm. This extended color map enables a comprehensive view of the nonlinear response landscape and highlights the emergence of power-sensitive features.

As described by the four key features in Table~\ref{tbl:FourKeyFeatures}, nontrivial $P_\mathrm{3f}(T)$ structures arising from rf vortex nucleation by surface defects tend to shift to lower temperatures as the rf field amplitude increases. In a $P_\mathrm{3f}$ color map such as Fig.~\ref{fig:SnplatedTPsweep}, where temperature is plotted on the horizontal axis and input power on the vertical axis, this behavior manifests as a structure with a negative slope.

The white dashed box in Fig.~\ref{fig:SnplatedTPsweep} highlights the region containing the three nontrivial $P_\mathrm{3f}$ structures observed between 14~K and 16~K, which are shown in greater detail in Fig.~\ref{fig:SnplatedTsweepmultipleP}. In addition to these three structures, another distinct $P_\mathrm{3f}$ structure appears in the lower-left corner of Fig.~\ref{fig:SnplatedTPsweep}, within the range $-40~\mathrm{dBm} < P < -30~\mathrm{dBm}$ and $4~\mathrm{K} < T < 6~\mathrm{K}$. This $P_\mathrm{3f}$ structure, indicated by the orange and yellow region, has a transition temperature $T_\mathrm{c}^\mathrm{P_{3f}}$ near 6~K and exhibits a pronounced negative slope. Such negative slope behavior is consistent with the characteristic signatures of rf semi-loop vortex nucleation by surface defects. Taken together, all four $P_\mathrm{3f}$ structures display the key features outlined in Table~\ref{tbl:FourKeyFeatures}, supporting their interpretation as nonlinear responses originating from localized rf vortex nucleation at defect sites.

Compared to the lower-left $P_\mathrm{3f}$ structure, the three $P_\mathrm{3f}$ structures observed between 14~K and 16~K appear only at significantly higher input powers. This indicates that the corresponding rf vortices nucleate under stronger rf field amplitudes, suggesting higher local vortex penetration thresholds at those defect sites. Such behavior may result from one of two scenarios: (1) the defects possess intrinsically higher local vortex penetration fields, or (2) the defects are located deeper beneath the surface, or at some lateral distance from the probe, requiring stronger rf fields to activate their nonlinear response.

The strongest $P_\mathrm{3f}$ signal in Fig.~\ref{fig:SnplatedTPsweep}, represented by the large red region below 12 K, has a very weak negative slope trend. This behavior suggests that it may originate from a nonlinear mechanism distinct from rf semi-loop vortex nucleation at grain boundaries. At this point, we do not have a definitive microscopic explanation of this feature.


\section{Discussion}
\label{sec:Discussion}

Both Nb\textsubscript{3}Sn films studied in this work—the vapor-diffused (see Figs.~\ref{fig:vapordiffuseTsweepmultipleP} and \ref{fig:vapordiffuseTPsweep}) and electrochemical (see Fig.~\ref{fig:SnplatedTPsweep}) samples—exhibit nontrivial third-harmonic response structures below 7~K. These $P_\mathrm{3f}(T)$ structures occur well below the intrinsic superconducting transition temperature of stoichiometric Nb\textsubscript{3}Sn ($T_\mathrm{c} \approx 18$~K) and consistently display the four key characteristics associated with rf vortex nucleation by surface defects, as outlined in Table~\ref{tbl:FourKeyFeatures}. The observation of similar $P_\mathrm{3f}$ structures in both films, despite their different deposition methods and surface properties, suggests that they likely share comparable types of surface defects capable of nucleating rf vortices. 

It is known that a number of second phases  can exist in Nb\textsubscript{3}Sn films.  For example two other superconducting inter-metallic compounds exist in the Nb-Sn system, and have reported transition temperatures of $T_c=2.68$ K (NbSn$_2$) and $T_c=2.07$ K (Nb$_6$Sn$_5$), \cite{Charlesworth66} both well below our measurement temperature range.  Some Nb\textsubscript{3}Sn films are known to have residual Sn droplets on the surface \cite{SayeedSnDrop19,JiangSnDroplet24}.  In some vapor-diffused Nb\textsubscript{3}Sn samples, a surface Sn layer overlies the native oxides, giving a local cross-sectional sequence of Sn / oxides / Nb\textsubscript{3}Sn / Nb \cite{BarComp24}. 
However, the superconducting transition temperature of pure Sn is known to be T$_c = 3.72$ K \cite{CorakSn56}, just at the lower edge of our measurement range.  It should be noted that no evidence of a transition near the T$_c$ of the Nb substrates is seen in either film.

On the other hand, the two Nb\textsubscript{3}Sn films do exhibit notable differences in the number, temperature distribution, and power thresholds of the observed $P_\mathrm{3f}$ structures that are likely associated with rf vortex nucleation by surface defects. The vapor-diffused film displays two distinct $P_\mathrm{3f}$ structures, both appearing below 7~K, whereas the electrochemical film exhibits four such structures—one below 6~K and three between 14~K and 16~K. In contrast, no nontrivial $P_\mathrm{3f}$ structures are observed above 7~K in the vapor-diffused film. The three $P_\mathrm{3f}$ structures between 14~K and 16~K in the electrochemical film appear only at relatively high rf field amplitudes. At lower rf power levels—for example, $P = -30~\mathrm{dBm}$—these defects likely do not nucleate rf vortices in our local measurements.

The observation of $P_\mathrm{3f}$ structures with $T_\mathrm{c}^\mathrm{P_{3f}}$ below the intrinsic $T_\mathrm{c}$ of stoichiometric Nb\textsubscript{3}Sn ($\approx$ 18~K) suggests that certain surface regions of the films exhibit degraded superconducting properties. A likely origin of this behavior is local non-stoichiometry, particularly Sn deficiency, which is known to significantly suppress $T_\mathrm{c}$ \cite{Moore1979Energy,Devantay1981Physical,Li2017Glag,Sitaraman2021Effect,Sun2021Toward,Sun2023Smooth}. Notably, $T_\mathrm{c}$ drops to approximately 7 (5)~K for Sn concentrations near 18 (14)~\% \cite{Moore1979Energy}, consistent with the $P_\mathrm{3f}$ transition temperatures observed in this work. In vapor-diffused Nb\textsubscript{3}Sn, spatially nonuniform Sn vapor supply during growth can result in Sn-deficient grains and stoichiometric inhomogeneity near the surface. It is also known that the vapor-diffused Nb\textsubscript{3}Sn growing from the Nb interface often suffers from Sn deficiency, and the film thickness can vary considerably over the sample (for example, see the cross-section Sn composition images in Fig. 6 of Ref.~\cite{Sun2023Smooth}).  Although the electrochemical synthesis method improves average stoichiometric uniformity, isolated Sn-deficient regions have also been reported near the surface \cite{Sun2023Smooth}. These Sn-deficient regions likely give rise to the observed low-temperature $P_\mathrm{3f}$ structures at 6.5 K and below in both films.  It has also been calculated that strongly overdoped A15 Nb-Sn films will show transition temperatures as low as 5 K at 31\% Sn \cite{Sitaraman2021Effect}.


\section{Conclusion}
\label{sec:Conclusion}

In this work, we use our near-field magnetic microwave microscope to study rf vortex nucleation in two SRF-quality Nb\textsubscript{3}Sn films prepared by different fabrication methods, linking the response to surface properties. Both the vapor-diffused and electrochemical films exhibit nontrivial third-harmonic response structures below 7~K. The observation of similar low-temperature nonlinear features in both films suggests that they likely share comparable types of surface defects, despite differences in their growth techniques. In addition, the electrochemical film displays additional $P_\mathrm{3f}$ structures between 14~K and 16~K that are absent in the vapor-diffused film, indicating variations in the distribution or characteristics of these defects. These results suggest that fabrication methods can influence not only surface roughness and stoichiometry, but also the rf vortex penetration landscape. This measurement technique provides a powerful diagnostic tool for identifying surface defects and guiding the optimization of Nb\textsubscript{3}Sn films for SRF applications.


\section{Acknowledgement}
\label{sec:Acknowledgement}

The authors would like to thank Javier Guzman from Seagate Technology for providing magnetic write probes, and Nathan Sitaraman for helpful discussions. This work is funded by the U.S. Department of Energy/High Energy Physics through grant No. DESC0017931, the Army Research Office under Grant Number ARO W911NF-24-1-0153, and the Maryland Quantum Materials Center.

\appendix


\section{Additional Data Collected After Surface Damage to the Electrochemical Nb\textsubscript{3}Sn Film}
\label{sec:DamagedResult}

This appendix presents $P_\mathrm{3f}$ measurements performed on the electrochemical Nb\textsubscript{3}Sn film after the creation of a visible surface scar, caused by unintentional mechanical contact and deformation with the microscope probe (see Fig.~\ref{fig:samplephoto}(b)). These measurements were collected during a later experimental session and are not included in the main analysis. They are presented here to highlight the distinct nonlinear response that emerged following the damage.

\begin{figure}
\includegraphics[width=0.45\textwidth]{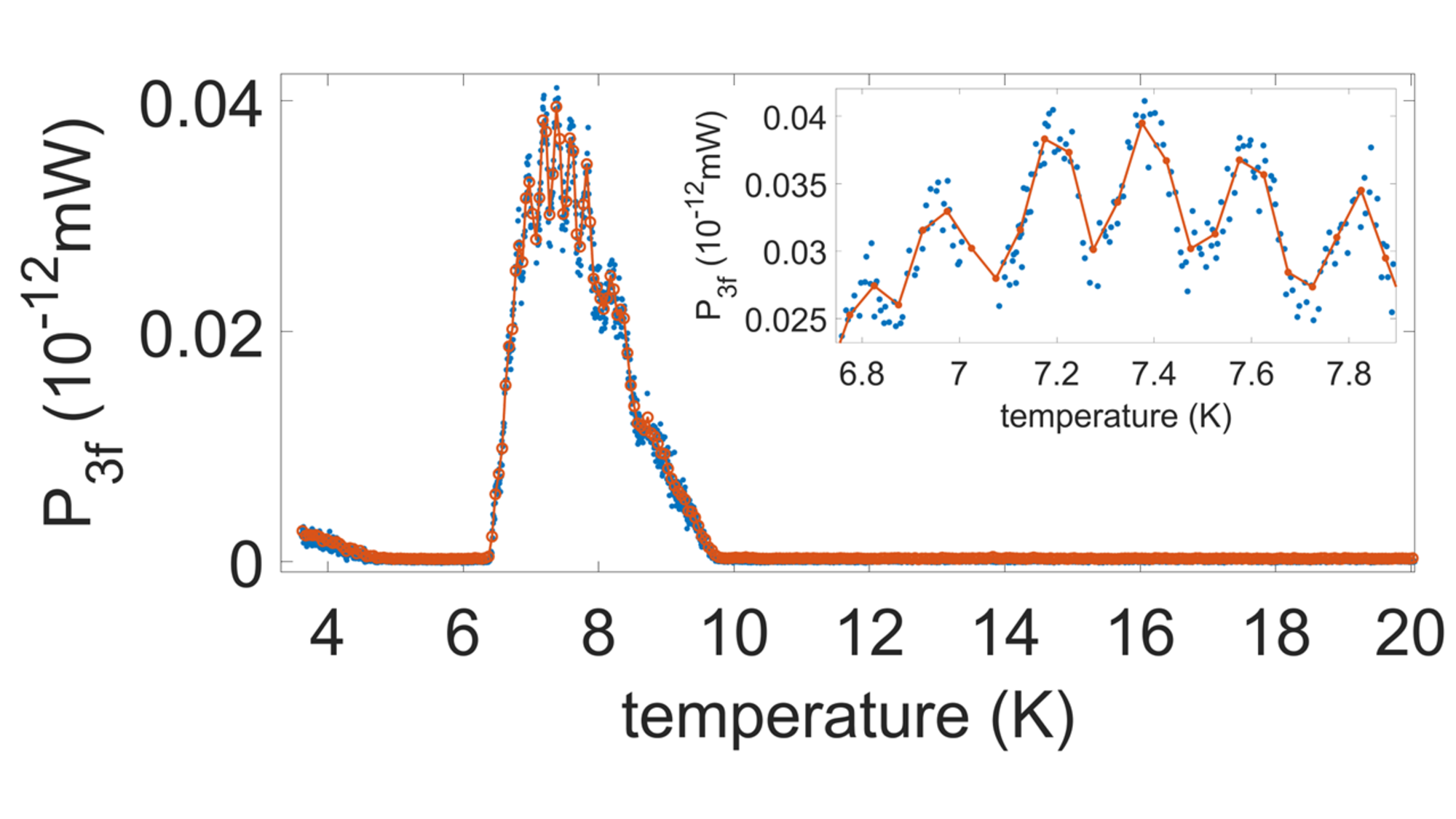}
\caption{\label{fig:SnplatedafterdamagedTsweep}Third-harmonic response $P_\mathrm{3f}$ as a function of temperature for the electrochemical Nb\textsubscript{3}Sn film after surface damage. The input frequency is 1.84~GHz and the input power is $-14$~dBm. The probe background has been subtracted. The inset shows a magnified view of the $P_\mathrm{3f}(T)$ structure between 6.8~K and 7.8~K.}
\end{figure}

Figure~\ref{fig:SnplatedafterdamagedTsweep} shows the third-harmonic response $P_\mathrm{3f}(T)$ measured on the electrochemical Nb\textsubscript{3}Sn film after surface damage. The measurement was performed at an input frequency of 1.84~GHz and an input power of $-14$~dBm, with the probe background subtracted. A notable enhancement in $P_\mathrm{3f}$ is observed below 9.8~K. The inset highlights an oscillatory structure in $P_\mathrm{3f}(T)$ between 6.8~K and 7.8~K, which is not present in the pre-damage measurements (see Figs.~\ref{fig:SnplatedTsweepmultipleP} and \ref{fig:SnplatedTPsweep}). This feature may reflect modified local electrodynamic properties introduced by the surface scar.

\begin{figure}
\includegraphics[width=0.45\textwidth]{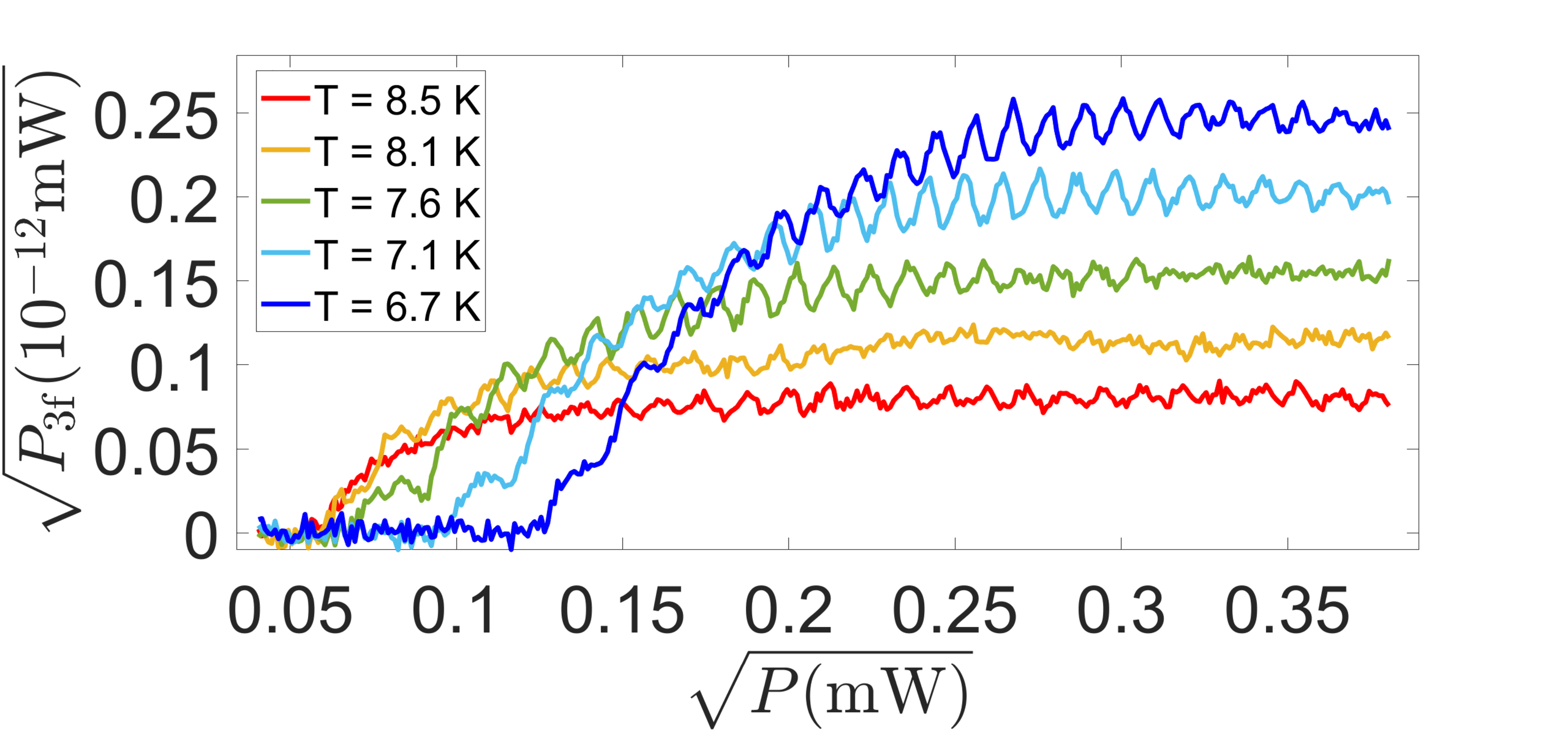}
\caption{\label{fig:SnplatedafterdamagedPsweepmultipleT}Third-harmonic response $\sqrt{P_\mathrm{3f}}$ as a function of the square root of input power $\sqrt{P}$ at five different temperatures for the electrochemical Nb\textsubscript{3}Sn film after surface damage. The measurement was performed at a fixed input frequency of 1.84~GHz, and the probe background has been subtracted. Note that $\sqrt{P}$ is proportional to the applied rf magnetic field amplitude.}
\end{figure}

The oscillatory behavior is also evident in the rf field amplitude dependence. Figure~\ref{fig:SnplatedafterdamagedPsweepmultipleT} presents $\sqrt{P_\mathrm{3f}}$ as a function of $\sqrt{P}$ at five temperatures ranging from 6.7~K to 8.5~K. The measurements were taken at an input frequency of 1.84~GHz, with the probe background subtracted. Since $\sqrt{P}$ is proportional to the applied rf magnetic field amplitude, these curves reflect the nonlinear response to increasing rf field amplitude. Pronounced oscillations in $\sqrt{P_\mathrm{3f}}$ are observed across all temperatures, with the effect becoming more pronounced as temperature decreases. The oscillations in $\sqrt{P_\mathrm{3f}}$ are approximately periodic as a function of $\sqrt{P}$ on a linear scale. For reference, an input power of $P = -14$~dBm corresponds to $\sqrt{P~(\mathrm{mW})} = 0.2$.

The oscillatory patterns observed in $P_\mathrm{3f}(T)$ and $P_\mathrm{3f}(\sqrt{P})$ (Figs.~\ref{fig:SnplatedafterdamagedTsweep} and \ref{fig:SnplatedafterdamagedPsweepmultipleT}, respectively) are absent in the pre-damage measurements (Figs.~\ref{fig:SnplatedTsweepmultipleP} and \ref{fig:SnplatedTPsweep}) and are likely associated with changes to the surface introduced by the scar. Similar oscillatory behavior has been reported in a previous study of SRF-grade Nb samples~\cite{oripov2019high}, where it was attributed to the rf response of weak-link Josephson junctions. In the present case, it is possible that pressing the microscope probe against the electrochemical Nb\textsubscript{3}Sn film with a substantial force modified the film surface and locally formed weak-link regions beneath the microscope probe, which could explain the emergence of the oscillatory patterns in the third-harmonic response. These results underscore the sensitivity of third-harmonic measurements to local surface conditions and highlight their potential as a diagnostic for identifying weak-link behavior induced by mechanical or structural imperfections.


\bibliography{Nb3SnReference.bib}

\end{document}